\documentstyle[12pt]{article}

\newcommand{\rbox}[1]{\vbox{\hrule height.8pt%
                \hbox{\vrule width.8pt\kern5pt
                \vbox{\kern5pt\hbox{#1}\kern5pt}\kern5pt
                \vrule width.8pt}
                \hrule height.8pt}}

\begin{document}


\newif\iffigureexists
\newif\ifepsfloaded
\openin 1 epsf.sty
\ifeof 1 \epsfloadedfalse \else \epsfloadedtrue \fi
\closein 1
\ifepsfloaded
    \input epsf.sty
\else
    \immediate\write20{>Warning:
         No epsf.sty --- cannot embed Figures!!}
\fi
\def\checkex#1 {\relax
    \ifepsfloaded \openin 1 #1
        \ifeof 1 \figureexistsfalse
        \else \figureexiststrue
        \fi \closein 1
    \else \figureexistsfalse
    \fi }

\def\epsfhako#1#2#3#4#5#6{
\checkex{#1}
\iffigureexists
    \begin{figure}[#2]
    \epsfxsize=#3
    \centerline{\epsffile{#1}}
    {#6}
    \caption{#4}
    \label{#5}
    \end{figure}
\else
    \begin{figure}[#2]
    \caption{#4}
    \label{#5}
    \end{figure}
    \immediate\write20{>Warning:
         Cannot embed a Figure (#1)!!}
\fi
}

\ifepsfloaded\baselineskip=18pt plus 0.2pt minus 0.1pt
\checkex{graph12.eps}
    \iffigureexists \else
    \immediate\write20{>EPS files for Figs. 2 and 3 are packed
     in a uuecoded compressed tar file}
    \immediate\write20{>appended to this LaTeX file.}
    \immediate\write20{>You should unpack them and LaTeX again!!}
    \fi
\fi

\renewcommand{\thepage}{}

\begin{titlepage}
\title{
\hfill
\parbox{4cm}{\normalsize KUNS-1298\\HE(TH)\ 94/14 \\
hep-ph/9411201}\\
\vspace{5ex}
Possible Candidates for SUSY $SO(10)$ Model\\
with an Intermediate Scale
\vspace{5ex}}

\author{Masako Bando
\thanks{e-mail address: {\tt mband@jpnyitp.bitnet}}\\
  {\it Physics Division, Aichi University,
       Aichi 470-02, Japan}\vspace{2ex}\\
{\rm Joe Sato
   \thanks{e-mail address: {\tt joe@gauge.scphys.kyoto-u.ac.jp}}
   and Tomohiko Takahashi
\thanks{e-mail address: {\tt tomo@gauge.scphys.kyoto-u.ac.jp}}}\\
  {\it Department of Physics, Kyoto University,
      Kyoto 606-01, Japan}}
\date{\today}
\maketitle

\begin{abstract}
\normalsize
We study the possibility of an intermediate scale existing in
supersymmetric $SO(10)$ grand unified theories: The intermediate scale
is demanded to be around $10^{12}$ GeV so that
neutrinos can obtain masses
suitable for explaining the
experimental data on the deficit
of solar neutrino with Mikheev-Smirnov-Wolfenstein solution and the existence
of
hot dark matter.
We show that any Pati-Salam type intermediate symmetries are
excluded by requiring reasonable conditions and
only $SU(2)_L\times SU(2)_R \times SU(3)_C\times U(1)_{B-L}$
is likely to be realized as an intermediate symmetry.
\end{abstract}
\end{titlepage}

\newpage
\renewcommand{\thepage}{\arabic{page}}
\setcounter{page}{2}

In constructing a realistic unified theory of matters and fields,
it is inevitable to answer a question about neutrino masses.
There seems to exist experiments indicating the neutrino
masses and their mixing \cite{yana}:
some experiments show
a deficit of the solar neutrino, which may be explained
by Mikheev-Smirnov-Wolfenstein (MSW) solution \cite{MSW}.
For example,
according to one of MSW solutions,
the mass of muon neutrino seems to be
$m_{ \nu_\mu}\simeq10^{-3}$ GeV.
Those small masses may be explained by the see-saw
mechanism \cite{seesaw}: muon neutrino can
acquire such a small mass
if Majorana mass of right-handed muon neutrino
is about $10^{12}$ GeV.
Furthermore
if all Majorana masses of right-handed neutrinos
are of about $10^{12}$ GeV,
the see-saw mechanism leads to the mass of tau neutrino
$m_{\nu_\tau} \simeq 10$ eV, which is desirable for the
interpretation that tau neutrino may be hot dark matter.
In the framework of $SO(10)$ Grand Unified Theory (GUT) \cite{so10},
we can naturally incorporate right-handed neutrinos
into ordinary fermions.

On the other hand it is well known that in Minimal Supersymmetric
Standard Model (MSSM) the present experimental values
of gauge couplings are successfully unified at a unification scale
$M_U \simeq 10^{16}$GeV \cite{Amal}.

Then how can the right-handed neutrinos
acquire masses of  about $10^{12}$ GeV
when we have no scale other than $M_U$?
There are
several possibilities for the right-handed neutrinos to obtain
masses of the intermediate scale, $M_R\simeq10^{12}$ GeV.
First, radiative
correction of GUT scale physics, what we call Witten mechanism \cite{witten},
can induce $M_R$.
In a supersymmetric model, however, this
mechanism cannot work because the
non-renormalization theorem \cite{nonreno} does
protect inducement of terms via radiative corrections which are not contained
in an original
Lagrangian. The second possibility is that the Yukawa coupling of right-handed
Majorana neutrino is so small that the mass may be the
intermediate scale even if it originates in GUT scale. Thirdly singlet
Higgs particles develop the vacuum expectation value at the
intermediate scale mass to supply the mass of
$M_{R}$ to $\nu_R$.
In unrenormalizable models such as supergravity
those latter two possibilities may be realized.

Our point of view is, however, that it is more natural to consider
that one energy scale corresponds to a dynamical phenomenon,
for instance a symmetry breaking.
Thus we are led to another possibility that a certain group breaks
down to the standard group at the intermediate scale at which right-handed
neutrinos gain masses.
This idea is consistent with the survival hypothesis.
There are some
papers which indicate that intermediate groups can enter in
breaking chains of groups from GUT to Standard Model (SM)
consistently with the analyses of MSSM such as coupling unification
\cite{Kawa}, but models presented in
these papers do not
involve Higgs contents which are able to provide masses to
right-handed neutrinos. We would like to explain
right-handed neutrino masses according to a symmetry breaking.

This paper is devoted to an investigation of a possibility
that right-handed neutrinos acquire masses of order $10^{12}$GeV
through a symmetry breaking within SUSY $SO(10)$ GUT
with one intermediate scale
below which MSSM is realized.
First we show assumptions about models and constraints
on our analysis. Next we explain our analysis. Then we
show results. Finally we give a summary.

First of all we show the breaking patterns allowed in our scenario,
\begin{eqnarray}
&&
{\rm I.}\ \ \ SO(10) \longrightarrow SU(2)_L \times SU(2)_R \times SU(4)_{PS}
         ~(G_{224})\longrightarrow {\rm MSSM},
\nonumber\\
&&
{\rm II.}\ \ SO(10) \longrightarrow SU(2)_L \times SU(2)_R \times SU(3)_{C}
             \times U(1)_{B-L} ~(G_{2231})\longrightarrow {\rm MSSM},
\nonumber\\
&&
{\rm III.}\ SO(10) \longrightarrow SU(2)_L \times U(1)_R \times SU(4)_{PS}
           ~(G_{214})\longrightarrow {\rm MSSM}.
\end{eqnarray}

The first assumption is that there is at least one Higgs multiplet
which is able to supply the intermediate scale mass to right-handed
neutrinos besides ordinary matter (ordinary fermions, right-handed
neutrinos and two Higgs doublets).

The second is that candidates for the matter contents in the intermediate
region
are multiplets
included in representations {\bf 10, 16, 45, 54, 120, 126} and {\bf 210}
of $SO(10)$.\footnote{This is just an assumption. In general,
however, models mentioned above
are difficult to be accompanied by
a representation which is contained
in a higher representation of $SO(10)$ only. See a statement below eq.
(\ref{eq:Input}) and the summary for an example.}

Phenomenologically we impose further constraints to our
models.
\begin{enumerate}
\item The unified scale $M_U$ is larger than $10^{16}$ GeV. This is
necessary for suppression of proton decay \cite{proton}.
\item The intermediate scale is taken at $10^{10}$, $10^{11}$,
$10^{12}$, $10^{13}$ or $10^{14}$ GeV because of right-handed
neutrino masses. The reason why an intermediate scale
is an input will be made clear after eq.(4).
\item The unified gauge coupling $\alpha_U(M_U)$ satisfies
$\alpha^{-1}_U(M_U) \geq 5$. This constraint is required for
a perturbative region to expand to GUT scale.
\item Any colored Higgs is not contained in the intermediate physics. This
is needed also for suppression of proton decay \cite{proton}.
\end{enumerate}
 Under these conditions, we specify combinations of matters
mentioned above which realize the unification of gauge
couplings to achieve GUT with a simple group.

Next we show the condition on the beta functions in the intermediate
scale in order to achieve the unification of gauge couplings.
In the following we made an analysis based on RGE up to one loop.
The conditions of the unification are described by
\begin{eqnarray}
\alpha^{-1}_{Y}(M_S) &=& \alpha^{-1}_{U}(M_U) + \frac{1}{2\pi} b_Y R
                            + \frac{1}{2\pi} b'_Y (U - R),
\nonumber\\
\alpha^{-1}_{L}(M_S) &=& \alpha^{-1}_{U}(M_U) + \frac{1}{2\pi} b_{L} R
                            + \frac{1}{2\pi} b'_{L} (U - R),
\\
\alpha^{-1}_{C}(M_S) &=& \alpha^{-1}_{U}(M_U) + \frac{1}{2\pi} b_{C} R
                            + \frac{1}{2\pi} b'_{C} (U - R),
\nonumber
\end{eqnarray}
$b_i (i=Y,L,C)$'s with dash and without dash denote the
beta function in the lower scale and higher scale than the
intermediate scale $M_R$, respectively.
R and U are defined by
\begin{eqnarray}
R={\rm ln} \frac{M_R}{M_S},
\hspace{3em}
U={\rm ln} \frac{M_U}{M_S}.
\end{eqnarray}
These equations lead to the relation which R and U must satisfy,
\begin{eqnarray}
(b_Y - b_{L}) R + (b'_Y - b'_{L}) (U-R)
&=&
2\pi\left(\alpha^{-1}_{Y}(M_S) - \alpha^{-1}_{L}(M_S)\right),
\nonumber\\
(b_{C} - b_{L}) R + (b'_{C} - b'_{L}) (U-R)
&=&
2\pi\left(\alpha^{-1}_{C}(M_S) - \alpha^{-1}_{L}(M_S)\right).
\label{eq:ur}
\end{eqnarray}
Here we have assumed that in the lower scale MSSM is realized,
so the equation (\ref{eq:ur}) has always a solution  $U=R \simeq
10^{16}$ GeV, which corresponds
to the case that there is  no intermediate scale physics.
Therefore if there is a nontrivial
intermediate scale $R$, the beta
functions must satisfy the following condition,
\begin{eqnarray}
(b_Y - b_{L})(b'_{C} - b'_{L}) -
(b_{C} - b_{L})(b'_Y - b'_{L}) = 0.
\end{eqnarray}
Since the beta functions in MSSM are given by
\begin{eqnarray}
b_Y = \frac{33}{5},\ \ \ b_{L} = 1,\ \ \ b_{C} = -3,
\end{eqnarray}
the beta functions between the intermediate scale and GUT
scale must satisfy the equation,
\begin{eqnarray}
5 b'_Y - 12 b'_{L} + 7 b'_{C} = 0. \label{eq:UC}
\end{eqnarray}
which we call ``the unification condition''.\footnote{Though $b'_Y = b'_L
= b'_C$ satisfies the
unification condition, in this case the condition that
all couplings are unified is not fulfilled. Therefore this
case is excluded.}
This is a sufficient condition on the gauge coupling unification
under the assumption that MSSM is realized in the lower scale.
When the equation (\ref{eq:UC}) is satisfied, R becomes an arbitrary
parameter. Therefore we introduce an intermediate scale.

We make an analysis as follows using the unification condition for the
beta functions in addition to the above restrictions. Taking one
combination of matter
contents on the intermediate physics, we see whether the unification
condition is fulfilled or not. If it is the case,
we can calculate the unified scale $M_U$
and the gauge coupling $\alpha_U(M_U)$
at the unified scale using following equations,
\begin{eqnarray}
M_U \hspace{1em}&=& M_R\, {\rm exp}\left(2 \pi \frac{\alpha^{-1}_Y
                       (M_R)-\alpha^{-1}_{L}(M_R)}
                        {b'_Y - b'_{L}}\right),
\nonumber\\
\alpha_U(M_U) &=& \left(\alpha^{-1}_{L}(M_R) - \frac{1}{2\pi}
                b'_{L}(U-R) \right)^{-1},
\end{eqnarray}
once $M_R$ and $\alpha^{-1}_i(M_R)$'s are given.

In principle we can calculate $\alpha^{-1}_i(M_R)$'s from low-energy
experimental values of  $\alpha^{-1}_i$'s according to the RGE.
We choose, however, another
way to calculate  $\alpha^{-1}_i(M_R)$ in order
to avoid ambiguities such as SUSY breaking scale
$M_S$, strong coupling $\alpha_{C}$ and so on.
Because we already know the unification scale $M_U^{\rm MSSM}$ and
${\alpha^{\rm MSSM}}^{-1}_U(M_U)$ in MSSM GUT and above the intermediate
scale considered in this paper all couplings  $\alpha_i$'s
are small enough for one-loop approximation of RGE to work well,
we calculate $\alpha^{-1}_i(M_R)$'s from the
input parameter ${\alpha^{\rm MSSM}}^{-1}_U(M_U)$ at
the GUT scale $M_U^{\rm MSSM}$.
We choose input parameters from ref.\cite{Amal} as follows,
\begin{eqnarray}
M_U^{\rm MSSM} = 10^{16.3}\, {\rm GeV},
\ \ {\alpha^{\rm MSSM}}^ {-1}_U(M_U)=25.7. \label{eq:Input}
\end{eqnarray}

Then we select the matter
contents which satisfy the criteria. A search was made for all
possible combinations of matter contents. The reason why this is
possible is that the number of each matter multiplet
which we can take into account simultaneously is limited.
If the number of a representation is sufficiently large,
its contribution to beta functions is too big to keep the couplings
$\alpha_i$'s in the perturbative region below the unification scale $M_U$
(see constraint 2).

Now we present our results. The case II. is the most
preferable group for
us accepting it as the intermediate one. In this case,
though many combinations of
matter multiplets satisfy the conditions which we imposed,
most combinations lead the gauge coupling $\alpha_U$
to the value of about 1/5, which is beyond the acceptable region to
apply the perturbation theory.

There are only two combinations which
result in small value of $\alpha_U \simeq 1/20$.
\begin{eqnarray}
{\rm Solution\ (i)}
&&\hspace{2em}
\alpha^{-1}_U(M_U) = 19.0049,\ \ M_U=10^{16.3}\ {\rm GeV}
\hspace{5em}\nonumber\\
&& \underline{{\rm Higgs\ contents}}
\nonumber\\
&&\hspace{2em}
\begin{array}{ccccc}
(1,3,1)(6) & 1 && (1,3,1)(-6) & 1 \\
(2,2,1)(0) & 2 &&&\\
(3,1,1)(0) & 1 &&&\\
(1,1,8)(0) & 1 &&&
\end{array}
\nonumber\\
&&\nonumber\\
{\rm Solution\ (ii)}&&\hspace{2em}
\alpha^{-1}_U(M_U) = 19.0049,\ \ M_U=10^{16.3}\ {\rm GeV}
\nonumber\\
&& \underline{{\rm Higgs\ contents}}
\nonumber\\
&&\hspace{2em}
\begin{array}{ccccc}
(1,3,1)(6) & 1 && (1,3,1)(-6) & 1 \\
(2,2,1)(0) & 1 &&&\\
(2,1,1)(3) & 1 && (2,1,1)(-3) & 1 \\
(3,1,1)(0) & 1 &&&\\
(1,1,8)(0) & 1 &&&
\end{array}
\end{eqnarray}
with an input parameter $M_R=10^{12}$GeV. In case of another $M_R$,
$\alpha^{-1}_U(M_U)$ is slightly varied, though $M_U$ does not change.
In this table,\footnote{We adopt the
normalization for $U(1)_{B-L}$  $T^{15}_4 = {\rm diag}(1,1,1,-3)$.} for
example, (1,3,1)(6) 1 stands
for that the representation
of the Higgs under $G_{2231}$ is (1,3,1)(6) and its number is one.
In the solution (i), (1,3,1)($\pm$6) are contained in {\bf 126} of
$SO(10)$ and possibly give Majorana mass to right-handed
neutrino. (2,2,1)(0) can be regarded as the standard Higgs in MSSM and
belonging to {\bf 10} or {\bf 126} of $SO(10)$. (3,1,1)(0) and
(1,1,8)(0) are involved in {\bf 45} of $SO(10)$ \cite{slan}.
The unification of the gauge couplings of the solution (i) is represented
in Figure \ref{fig:gr12}.

In contrast to the fact that many viable combinations exist in the case II,
there is no solution for the breaking chain of I, in which
$SO(10)$ breaks down to SM through so-called Pati-Salam
symmetry \cite{PatiSalam}.
Therefore if we consider both supersymmetric GUT and the scenario described
above, in which right-handed neutrino masses are
given by the symmetry breaking at the
intermediate scale, it is impossible to use
the group $G_{224}$ as the intermediate group.
In the breaking chain of III, there is no solution which passes all of the
constraints except the case of $M_R=10^{14}$GeV,\footnote{If ignoring
the constraint 4, there are solutions in the case
that  $M_R\ge 10^{12}$GeV. Proton decay via colored Higgs is quite
dependent on a model.} which is a rather high scale as
an intermediate scale.
Moreover in this case $\alpha_U$ becomes larger than 0.1 at the GUT
scale and we can not trust perturbative approach to the unified
scale. So it is difficult for the group
$G_{214}$ to be realized at high energy region in this scenario.

Finally we give a summary.
When we construct SUSY $SO(10)$ GUT model under the assumption
that the right-handed neutrinos acquire their masses of the intermediate
scale by the renormalizable coupling (Yukawa coupling)
and there is only one intermediate scale,
among three cases $G_{2231}$ is the most favorable group to be built
in. The reason why neither $G_{224}$ nor $G_{214}$
can be used is as follows.
In case I, to give right-handed neutrinos masses,
(1,3,10) Higgs is needed. This representation, however, makes
too large a contribution to beta functions to achieve a small coupling.

We thank T.~Kugo, K.~Inoue and Izawa,~K-I. for useful discussions related to
this work. M.B. is supported in part by Grant-in-Aid for Scientific
Research from
Ministry of Education, Science, and Culture(\#06640416). T.T. is
JSPS Research Fellow and is supported in part by Grant-in-Aid for Scientific
Research from Ministry of Education, Science, and Culture(\#3116).
\vspace{-2ex}

\epsfhako{graph12.eps}{thb}{32em}{Evolution of coupling constants under
the breaking chain given in II and the Higgs contents in solution
(i). $M_R$ is taken to be
$10^{12}$GeV. The calculation is  based on one-loop approximation.}{fig:gr12}{}

\newpage
\newcommand{\J}[4]{{\sl #1} {\bf #2} (19#3) #4}

\end{document}